\documentclass[10pt,prd,twocolumn]{revtex4-1}
\usepackage{hyperref,graphicx,color,amsmath, amssymb,natbib}
\def\apj{ApJ}%
\def\apjl{ApJ}%
\def\aap{A\&A}%
\def\mnras{MNRAS}%
\def\figurewidth{0.47\textwidth}
\begin{document}
\title{Revisiting the cosmological bias due to local gravitational redshifts}
\author{Zhiqi Huang$^{1}$}
\affiliation{${}^1$Canadian Institute for Theoretical Astrophysics, University of Toronto}

\date{\today}
\begin{abstract}
  A recent article by Wojtak {\it et al} (arXiv:1504.00718) pointed out that the local gravitational redshift, despite its smallness ($\sim 10^{-5}$), can have a noticeable ($\sim 1\%$) systematic effect on our cosmological parameter measurements. The authors studied a few extended cosmological models (nonflat $\Lambda$CDM, $w$CDM, and $w_0$-$w_a$CDM) with a mock supernova data set. We repeat this calculation and find that the $\sim 1\%$ biases are due to strong degeneracy between cosmological parameters. When cosmic microwave background (CMB) data are added to break the degeneracy, the biases due to local gravitational redshift are negligible ($\lesssim 0.1 \sigma$).
\end{abstract}
\maketitle

\section{Introduction}

The cosmological principle, that the Universe is homogeneous and isotropic on large scales, has been a successful theoretical basis for modern physical cosmology. In this idealized homogeneous and isotropic ``background Universe'' we can define a cosmic redshift that accounts for the redshift due to background expansion $z_{c} \equiv a_0/a - 1$, where $a$ is the scale factor inthe  Friedmann-Lema\^{i}tre-Robertson-Walker metric,
\begin{equation}
  ds^2 = -dt^2 + a(t)^2 \left[ \frac{dr^2}{1-kr^2} + r^2\left(d\theta^2+\sin^2\theta d\phi^2\right)\right], \label{eq:metric}
\end{equation}
and the subscript $0$ denotes the quantity today. The normalization of $a(t)$ is arbitrary and chosen to be $a_0 = 1$ in this work. Natural units $c=\hbar = 1$ are used throughout the paper.

We are interested in the cosmological parameters constrained by measurements of luminosity distances. In the idealized background Universe, the luminosity distance $d_L$ is a function of cosmic redshift $z_c$. Thus, the observational task is to measure the luminosity distance for objects at different redshifts and confront the data with the theoretical $d_L(z_c)$.

On cosmological scales, measuring luminosity distance is a challenging task. Ideally, for a standard candle whose luminosity is known as {\it a priori} knowledge, the luminosity distance can be determined with the inverse-square law for apparent brightness. In practice, the best cosmological candidate by far -- the type Ia supernova -- is not a perfect standard candle.  The environmental dependence of supernova luminosity~\cite{Sullivan10} and dust extinction~\cite{Totani99} are accounted for by introducing some nuisance parameters in the supernova light-curve fitting methods (see ~\cite{Conley08} and the references therein for a review). Other systematics such as second-order corrections from gravitational lensing and directional averages have also been discussed in the literature~\cite{BenDayan:2012ct, DenDayan:2013gc, Bonvin15}.

Redshift by its definition is a direct observable. To linear order the observed redshift $z_{\rm obs}$ contains three components: $z_{\rm obs} = z_c + z_{\rm pec} + z_{\rm g}$. The redshift due to peculiar velocity $z_{\rm pec}$ adds random scatter that averages out. The local gravitational redshift $z_g$ accounts for the difference between the gravitational potentials at the points of light reception and emission, which does not cancel if our local gravitational potential differs from the average potential in the supernova environments. If we split the supernova samples into different redshift bins and consider $z_g$ for each subset of supernovae, $z_g$ can be redshift dependent. Moreover, in a Universe with time-dependent gravitational potentials, $z_g$ also contains an integrated Sachs-Wolfe contribution that has redshift dependence. Simulations show, however, that $z_g$ can be approximately treated as a constant at $z\lesssim 1$ and its amplitude is $\le 4\times 10^{-5}$ at 95\% confidence level (C.L.)~\cite{Wojtak15}. This seemingly tiny systematic error by far has been ignored in supernova data analysis. Ref.~\cite{Wojtak15} for the first time computed the impact of a nonzero $z_g$ and found $\sim 1\%$ bias on the best-fit cosmological parameters. For modern precision cosmology, $1\%$ bias is a non-negligible effect. However, for the extended models (nonflat $\Lambda$CDM, $w$CDM, and $w_0$-$w_a$CDM) studied in Ref.~\cite{Wojtak15}, cosmological parameters are strongly degenerate with current supernova data {\it alone}. The $1\%$ shift of  best-fit parameters in the flat degeneracy direction may not be a significant effect. The purpose of this paper is to demonstrate that the systematic errors due to $z_g$ becomes negligible when we use cosmic microwave background (CMB) data to break the degeneracy between parameters.

\section{Systematic Errors with Supernova Data alone \label{sec:sn}}

The standard treatment of supernova likelihood is to absorb the random scatter $z_{\rm pec}$ into an additional uncertainty of luminosity distance. The local gravitational redshift is a biased error that, in principle, cannot be treated in the same way. With random scatter ignored, a nonzero $z_g$ gives an observed redshift
\begin{equation}
  z_{\rm obs} = z_{c} + z_g, \label{eq:zobs}
\end{equation}
and an observed luminosity distance
\begin{equation}
  d_{L, \rm obs} = d_L(z_c) (1+z_{\rm obs})/(1+z_c), \label{eq:dlobs}
\end{equation}
 where $d_L(z_c)$ is the theoretical luminosity distance function. (See Eq.~(3.3) of Ref.~\cite{Wojtak15}.) The extra factor $(1+z_{\rm obs})/(1+z_c)$, which has been ignored in Ref.~\cite{Wojtak15}, accounts for the fact that a local gravitational redshift also changes the apparent brightness via an additional redshift of photons and an additional time dilation. We find that, however, for all cases discussed in this paper the $(1+z_{\rm obs})/(1+z_c)$ correction does not lead to a noticeable difference.

 Following the approach in Ref.~\cite{Wojtak15}, we use the redshifts (treated as $z_c$) and errors in the apparent magnitude of 580 Union2.1 SN samples \cite{union2.1}. Mock data with a fiducial $\Lambda$CDM cosmology ($\Omega_m = 0.3$) and a fiducial $z_g$ are produced with Eqs.~(\ref{eq:zobs})-(\ref{eq:dlobs}). For mock data sets with $z_g = 0, \pm 4\times 10^{-5}$, we compute the best-fit parameters, respectively. The 68.3\%C.L. and 95.4\%C.L. constraints on parameters are computed via Monte Carlo Markov chain (MCMC) simulations.

 For illustration purposes, we consider a nonflat $\Lambda$CDM model and a flat $w$CDM model. (See \cite{Wojtak15} for the definitions of these models.) For supernova data the Hubble constant $H_0$ is degenerate with the supernova absolute luminosity and does not affect the likelihood. The likelihood depends on, in the nonflat $\Lambda$CDM case, $\Omega_k$ (curvature of the Universe) and $\Omega_m$ (fractional energy of matter), and in the $w$CDM case, $w$ (dark energy equation of state) and $\Omega_m$. The dark energy fractional energy $\Omega_\Lambda$ is constrained by $\Omega_k+\Omega_m +\Omega_\Lambda = 1$ ($\Omega_k = 0$ in $w$CDM case).

\begin{figure}
  \includegraphics[width=\figurewidth]{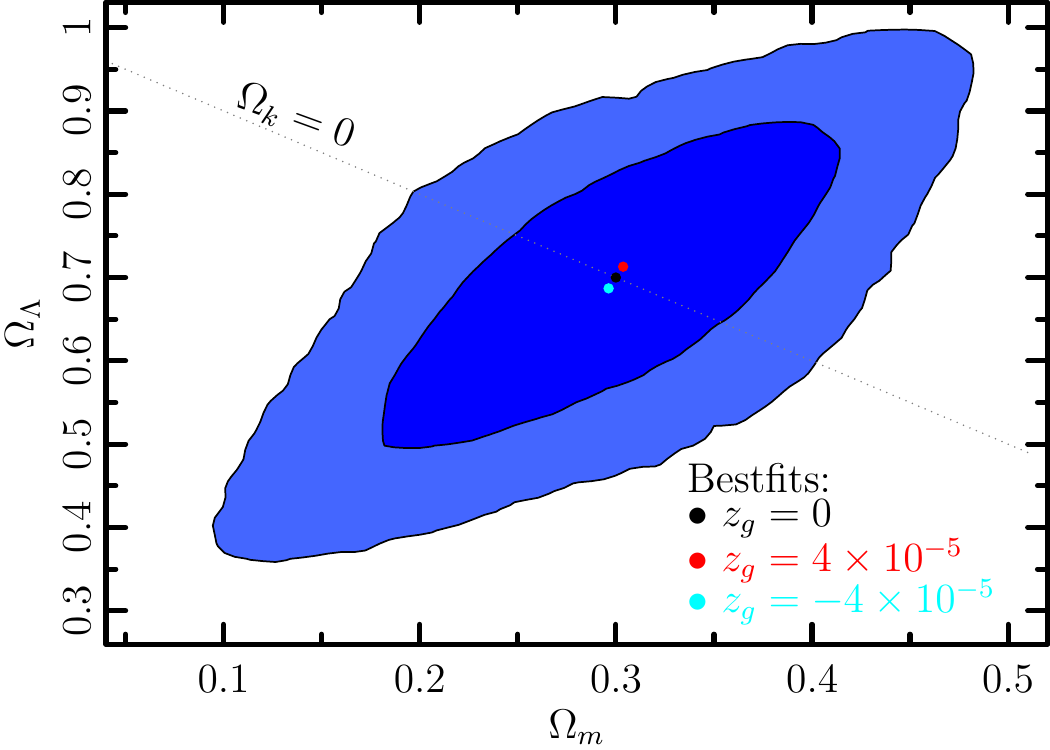}   
  \includegraphics[width=\figurewidth]{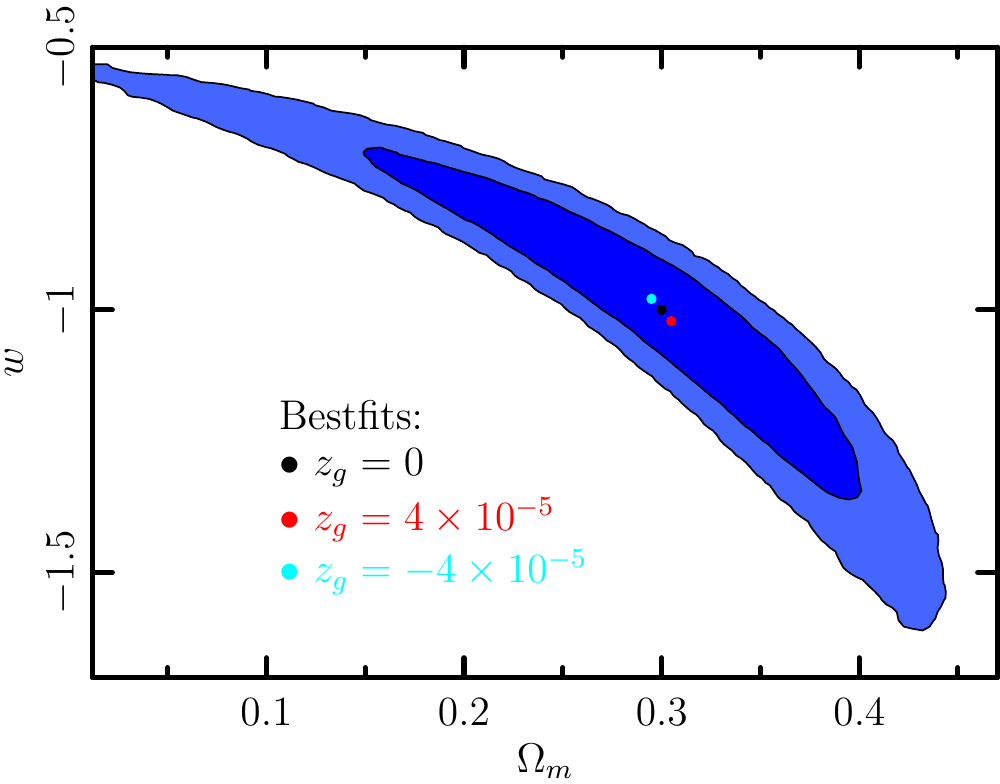}
  \caption{The systematic errors due to a local gravitational redshift. Upper panel: nonflat $\Lambda$CDM model; lower panel: flat $w$CDM model. The contours are 68.3\% C.L. and 95.4\% C.L. constraints for the $z_g=0$ case. \label{fig:sn}}
\end{figure}

As shown in Fig~\ref{fig:sn}, we find that the systematic errors on cosmological parameters agree with Ref.~\cite{Wojtak15}. However, the $\sim 1\%$ bias of best-fit parameters are aligned with the degeneracy direction where the posterior likelihood is flat. Compared to the $1\sigma$ (68.3\%~C.L.) constraint, the systematic errors are small.

\section{Systematic Errors with Supernova + CMB \label{sec:sncmb}}

In this section we combine the publicly available Planck CMB likelihood~\cite{pl13like} and the Union2.1 supernova likelihood to constrain cosmological models, with different fiducial $z_g$ assumed.  The Planck likelihood includes CamSpec temperature power spectrum likelihood (2013 release) and a low-$\ell$ WMAP polarization constraint. The supernova likelihood code is modified to include the $z_g$ correction.

We run Monte Carlo Markov chains with  the publicly available code COSMOMC~\cite{cosmomc}, with the standard settings, namely, flat priors on six parameters $\Omega_b h^2$ (baryon density), $\Omega_c h^2$ (cold dark matter density), $\theta$ (angular extension of sound horizon on the last scattering surface), $\tau$ (CMB optical depth), $\ln A_s$ and $n_s$ (logarithmic amplitude and index of the primordial power spectrum of curvature fluctuations). For nonflat $\Lambda$CDM model and $w$CDM model we add  $\Omega_k$ and $w$ with a flat prior, respectively. 

\begin{figure*}
  \includegraphics[width=\figurewidth]{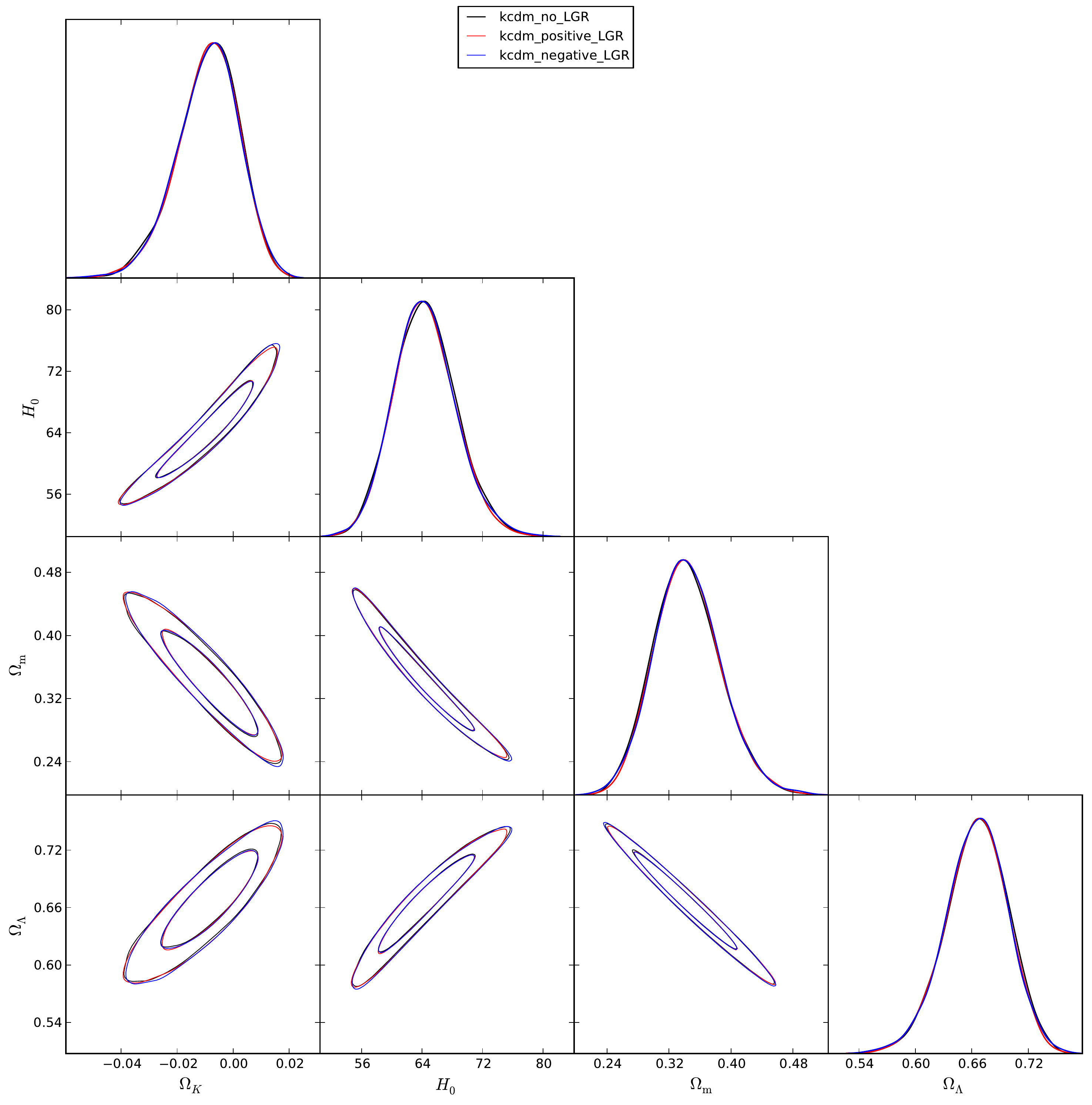}%
  \includegraphics[width=\figurewidth]{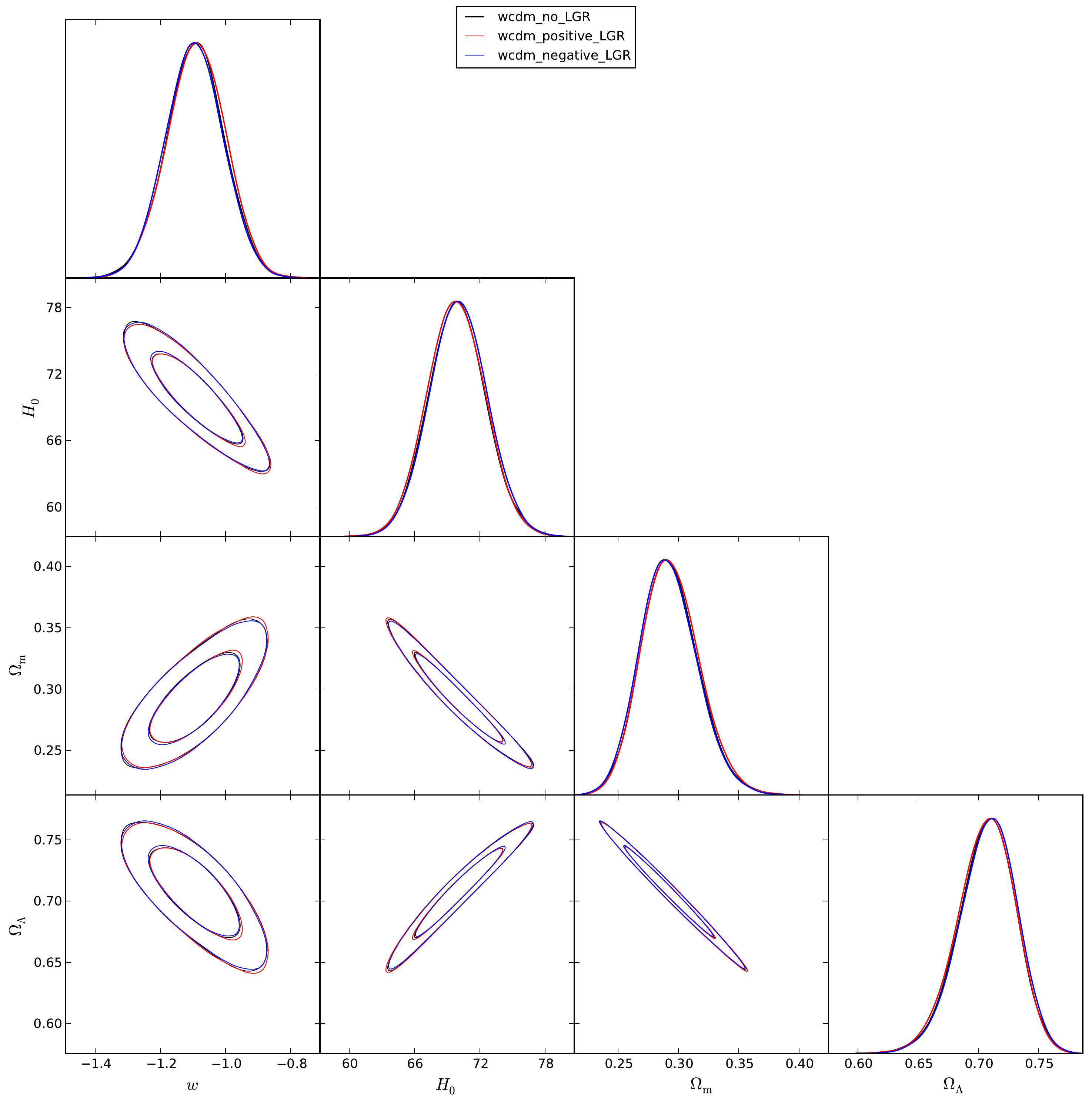}
  \caption{The 68.3\% C.L. and 95.4\% C.L. constraints on nonflat $\Lambda$CDM cosmology (left panel) and $w$CDM cosmology (right panel). The red, blue and black lines are for $z_g = 4\times 10^{-5}$, $z_g = 0$, and $z_g = -4\times 10^{-5}$ respectively. Data sets: Planck 2013 CamSpec +  WMAP low-$\ell$ polarization + Union2.1 Supernova. \label{fig:plsn}}
\end{figure*}

As shown in Fig.~\ref{fig:plsn}, the impact of $z_g$ becomes invisible with the addition of CMB data. In Table~\ref{tbl:kcdm} and Table~\ref{tbl:wcdm} we compare the mean and standard deviations of parameters and find that in all cases the biases due to a nonzero $z_g$ are less than $0.1\sigma$.

\begin{table}
    \caption{Nonflat $\Lambda$CDM; CMB + SN \label{tbl:kcdm}}  
  \begin{tabular}{llll}
    \hline
    \hline
    & $z_g = 0$ & $z_g = -4\times 10^{-5}$ & $z_g = 4\times 10^{-5}$  \\
    \hline
    $\Omega_m$ & $0.344 \pm 0.042$ & $0.344\pm 0.043$ & $0.345\pm 0.042$  \\
    $\Omega_\Lambda$ & $0.665 \pm 0.033$ &  $0.665\pm 0.032$ & $0.664\pm 0.032$ \\    
    $\Omega_k$ & $-0.009 \pm 0.011 $  &  $-0.009 \pm 0.011$ & $-0.009\pm 0.011$ \\
    \hline
  \end{tabular}
\end{table}

\begin{table}
  \caption{Flat $w$CDM; CMB + SN \label{tbl:wcdm}}  
  \begin{tabular}{llll}
    \hline
    \hline
    & $z_g = 0$ & $z_g = -4\times 10^{-5}$ & $z_g = 4\times 10^{-5}$  \\
    \hline
    $\Omega_m $ & $0.293\pm 0.023$ & $0.292\pm 0.023$ & $0.294\pm 0.023$ \\
    $w$ & $-1.093 \pm 0.087$ & $-1.096\pm 0.086$ & $-1.089\pm 0.087$ \\
    \hline
  \end{tabular}
\end{table}

\section{Conclusions and Discussions \label{sec:con}}

The findings of Ref.~\cite{Wojtak15}, that a local gravitational redshift can cause $\sim 1\%$ systematic errors, raise a question whether or not we should include the $z_g$ correction in the supernova data analysis. In this paper we have demonstrated that, for current supernova data, as long as CMB data are included to break the parameter degeneracy, the impact of the local gravitational redshift is negligible. Thus, no modification of current supernova likelihood is on demand.

In Sec.~\ref{sec:sn} and Sec.~\ref{sec:sncmb} we have used different sets of parameters. In particular, in the supernova-data-alone case we have used a flat prior on $\Omega_m$, and for CMB + SN we have used flat priors on $\Omega_b h^2$, $\Omega_c h^2$ and $\theta$. ($\Omega_m$ is a derived parameter with a complicated prior.) To  make a more direct comparison, we replace the CMB likelihood with a ``compressed CMB likelihood'' that contains only the distance information, that is, a Gaussian constraint $R = \sqrt{\Omega_m}H_0d_A(z_*) = 1.7488\pm 0.0074$ \cite{pl15DE}, where $d_A$ is the comoving angular diameter distance and $z_* = 1089$ is the redshift at recombination. The advantage of using this compressed likelihood is that $R$ can be written as a function of $\Omega_m$, $\Omega_k$ and $w$. For supernova and this compressed CMB likelihood, we use the same parametrization and priors as in Sec.~\ref{sec:sn} and find that the systematic errors due to $z_g$ remain negligible ($\lesssim 0.1\sigma$).

Whether a systematic error can be ignored depends on the precision of observations. Local gravitational redshift may still be a potential source of information contamination for future supernova observations. We leave the forecast of forthcoming supernova data to future work.


\begin{thebibliography}{1}

\bibitem{Sullivan10}
M.~{Sullivan} {\em et~al.},
\newblock \mnras {\bf 406}, 782 (2010), 1003.5119.

\bibitem{BenDayan:2012ct} 
  I.~Ben-Dayan, M.~Gasperini, G.~Marozzi, F.~Nugier and G.~Veneziano,
  Phys.\ Rev.\ Lett.\  {\bf 110}, no. 2, 021301 (2013)
  [arXiv:1207.1286 [astro-ph.CO]].
\bibitem{BenDayan:2013gc} 
  I.~Ben-Dayan, M.~Gasperini, G.~Marozzi, F.~Nugier and G.~Veneziano,
  JCAP {\bf 1306}, 002 (2013)
  [arXiv:1302.0740 [astro-ph.CO]].

\bibitem{Totani99}
T.~{Totani} and C.~{Kobayashi},
\newblock \apjl {\bf 526}, L65 (1999), astro-ph/9910038.

\bibitem{Conley08}
A.~{Conley} {\em et~al.},
\newblock \apj {\bf 681}, 482 (2008), 0803.3441.

\bibitem{Bonvin15}
C.~{Bonvin}, C.~{Clarkson}, R.~{Durrer}, R.~{Maartens}, and O.~{Umeh},
\newblock ArXiv e-prints  (2015), 1504.01676.

\bibitem{Wojtak15}
R.~{Wojtak}, T.~M. {Davis}, and J.~{Wiis},
\newblock ArXiv e-prints  (2015), 1504.00718.

\bibitem{union2.1}
N.~{Suzuki} {\em et~al.},
\newblock \apj {\bf 746}, 85 (2012), 1105.3470.

\bibitem{pl13like}
{Planck Collaboration} {\em et~al.},
\newblock \aap {\bf 571}, A15 (2014), 1303.5075.

\bibitem{cosmomc}
A.~{Lewis} and S.~{Bridle},
\newblock Phys. Rev. D {\bf 66}, 103511 (2002), arXiv:astro-ph/0205436.

\bibitem{pl15DE}
{Planck Collaboration} {\em et~al.},
\newblock ArXiv e-prints  (2015), 1502.01590.

\end{thebibliography}
\end{document}